\documentstyle[prb,aps,preprint,psfig]{revtex}
\tighten
\begin{document}
\draft 
\title{{\em Ab initio} Hartree-Fock Born effective
charges of LiH, LiF, LiCl, NaF, and NaCl} 
\author{Alok Shukla\cite{email}} 
\address{Department of Physics, University of Arizona,
Tucson, AZ 85721} 

\maketitle

\begin{abstract}
We use the Berry-phase-based theory of macroscopic polarization of dielectric
crystals formulated in terms of Wannier functions, and state-of-the-art
Gaussian basis functions, to obtain 
benchmark {\em ab initio} Hartree-Fock values of the 
Born effective charges of ionic compounds LiH, LiF, LiCl, NaF, and NaCl.
We find excellent agreement with the experimental values for  
all the compounds except LiCl and NaCl, for which the disagreement
with the experiments is close to 10\% and 16\%, respectively. 
This may imply the importance of many-body effects in those
systems.
\end{abstract}
\pacs{ }
\section{INTRODUCTION}
\label{intro}
The Born effective charge (also called transverse or dynamic charge) of
a crystalline system, defined as the induced polarization due to a unit
sublattice displacement~\cite{born}, is a fundamental quantity connecting
the electrostatic fields of the lattice, to its phononic properties.
It contains important information not only about the
electronic structure and the bonding properties of the system, but also
about the coupling of its longitudinal and transverse optical phonon modes to 
the external infrared radiation.~\cite{poster1,poster2,ghosez,bennetto,ogut}
In addition, the Born charges also find use in first-principles-based 
construction of effective Hamiltonians describing phase transitions in 
ferroelectric materials.~\cite{zvr,wagh}
Traditionally, the {\em ab initio} computations of  Born 
charges of dielectrics have been performed either 
within the density-functional linear-response theory,\cite{bald-res,wang-yu} 
or the density-functional perturbation theory.~\cite{ogut,testa}
At a more phenomenological level, the 
bond-orbital method of Harrison has been very insightful~\cite{harrison}.
Recently, however, a very elegant formalism has been proposed by
King-Smith and Vanderbilt,~\cite{king} and Resta,~\cite{resta-fe,resta} which 
formulates the general problem of symmetry breaking induced by the 
macroscopic polarization (of which the Born charge is a special case) of a 
crystalline dielectric, in terms of the Berry phase of its 
wave function. This Berry-phase-based approach to macroscopic polarization
of dielectrics has come to be known as the ``modern theory of 
polarization'' (MTP) in the current physics literature.~\cite{resta-rmp}
The MTP has been used both within the density-functional theory (DFT) based
implementations,~\cite{poster1,poster2,king,zhong,bern1,bern2,souza,%
parinello,car} as well as the wave-function-based 
Hartee-Fock (HF)~\cite{dall,yaschenko,fu} formulations, to evaluate a 
variety of  polarization related properties.

DFT-based calculations of  macroscopic polarization properties are 
 very efficient, so that, without excessive 
effort, one can perform {\em ab initio} computations on complex 
compounds. Normally, the {\em ab initio} Born charges computed using the 
DFT-based formulations are within 10\% agreement with the experiments for 
simple zinc-blende semiconductors,~\cite{bennetto} while the disagreement can
be worse for more complex systems.~\cite{resta-rmp} Therefore, it is of
interest to systematically explore alternative methods for computing the 
macroscopic polarization properties of crystalline insulators. For ionic 
systems, the HF method provides a powerful alternative in that it 
performs much better
than the local-density approximation based schemes.~\cite{fu,crystal}
The other advantage of the HF method is that it can be systematically
improved, both by perturbative, as well as nonperturbative methods, to
account for many-body effects.~\cite{fulde} Recently, we have developed
a Wannier-function-based {\em ab initio} HF approach to compute the 
ground-state properties of crystalline insulators.~\cite{shukla1,shukla2}
The approach has been successfully applied to compute the ground-state
properties of not only three-dimensional 
crystals,~\cite{albrecht,shukla3,shukla4,shukla5} but also of one-dimensional
periodic insulators such as polymers.~\cite{shukla6,shukla7,ayjamal}
In the present paper we extend our approach to the problem of macroscopic
polarization of dielectrics, and apply it to obtain benchmark HF values for 
the Born charges of diatomic ionic systems LiH, LiF, LiCl, NaF and NaCl. 
Besides their simplicity, the main criteria behind the choice of these 
materials for the present study were: (a) to the best of our knowledge, no 
prior benchmark calculations of the Born charges of these materials exist 
(b) high-quality experimental data  has been available for them for a 
long  time.~\cite{lucovsky,exp-lih,exp-ah} 
Thus by comparing the benchmark HF results such as this one,
to the corresponding experimental values, one can gauge the  
applicability of the HF method on a wide variety of systems.
When we  compare the HF values of the Born charges computed in the
present work, 
with the experimental ones, we find that the agreement is good for 
LiH, LiF, and NaF, where the agreement between the theory and the experiment
is always better than 7\%. For NaCl and LiCl, however,  the error is 10\% and 
16\%, respectively, suggesting the possibility that the
many-body effects may be stronger in these  systems.

Since, this is the first application of our Wannier-function-based method
to the problem of macroscopic polarization, we also present
the associated computational details.
The Wannier functions, being very similar in character to the 
molecular orbitals encountered routinely in quantum chemistry, have the
added advantage of being intuitive in character.
Indeed, as demonstrated later on, they lead to a pictorial
description of the symmetry breaking processes associated with macroscopic
polarization of insulators.

Remainder of the paper is organized as follows.  In section \ref{theory} 
we briefly cover the theoretical aspects of the present work.
Our numerical results for the Born effective charges of several ionic
crystals  are presented in section \ref{results}.  Finally, the  
conclusions are presented in section \ref{conclusion}.

\section{THEORETICAL BACKGROUND}
\label{theory}
\subsection{Born Effective Charge}   
\label{bec}
The Born effective charge tensor, $Z^*_{\alpha \beta}(i)$, associated
with the atoms of the $i$-th sublattice, is defined as~\cite{def-bec}
\begin{equation} 
Z^*_{\alpha \beta}(i) = 
Z_i + \left. (\Omega/e) \frac{\partial P^{(el)}_{\alpha}}
 {\partial u_{i \beta}} \right|_{{\bf E}={\bf 0}},
\label{zstar}
\end{equation}
where $Z_i$ is the charge associated with the nuclei (or the core) of 
the sublattice, $\Omega$ is the volume of the unit cell, $e$ is
the electronic charge, and
$P^{(el)}_{\alpha}$  is the $\alpha$-th Cartesian component of the
electronic part of the macroscopic polarization induced as a result
of the displacement of the sublattice in the $\beta$-th Cartesian
direction, $u_{i \beta}$.  
For small $\Delta u_{i \beta}$, 
one assumes $\left. \frac{\partial P^{(el)}_{\alpha}}
 {\partial u_{i \beta}} \right|_{{\bf E}={\bf 0}} = 
\frac{\Delta P^{(el)}_{\alpha}}{\Delta u_{i\beta}}$, and computes the
change in the polarization
$\Delta P^{(el)}_{\alpha}$ following Resta's approach~\cite{resta-fe}
\begin{equation}
 \Delta {\bf P}_{el} = {\bf P}^{(1)}_{el}-{\bf P}^{(0)}_{el} \; \mbox{,}
\label{delp-f}
\end{equation}
where, ${\bf P}^{(0)}_{el}$ and ${\bf P}^{(1)}_{el}$, respectively,
denote the electronic parts of the macroscopic polarization of the system
for its initial ($\lambda=0$) and final ($\lambda=1$) states, where $\lambda$
is a parameter characterizing the adiabatic symmetry-breaking transformation 
of the lattice. Clearly, for the present case, $\lambda$ is to be identified
with the sublattice displacement $\Delta u_{i \beta}$. For one-electron 
theories such as the Kohn-Sham theory, or the HF theory, King-Smith and 
Vanderbilt showed that~\cite{king}
\begin{equation}
 {\bf P}^{(\lambda)}_{el} = (fe/\Omega) \sum_{n=1}^{M} \int {\bf r}
\phi_n^{(\lambda)}({\bf r})^2 d{\bf r} \; \mbox{,} \label{p-wan}
\end{equation}
where \{ $\phi_n^{(\lambda)}({\bf r}), n=1,\ldots, M$ \} represent the $M$ 
occupied Wannier functions of the unit cell for a given value of $\lambda$,
and  $f$ is the occupation number of Wannier function ($f=2$, for the 
restricted-Hartree-Fock theory).
King-Smith and Vanderbilt~\cite{king} also showed that the r.h.s. of 
Eq. (\ref{p-wan}) is proportional to the
sum of the Berry phases associated with individual Wannier functions 
(or bands), thus equating the change in the macroscopic polarization of the
solid with the change in the Berry-phase of its wave function during the
corresponding adiabatic transformation. In addition, 
Resta~\cite{resta} demonstrated that $\Delta {\bf P}_{el}$ computed
via Eqs. (\ref{delp-f}) and (\ref{p-wan}) is invariant under the
choice of Wannier functions, even though the individual 
${\bf P}^{(\lambda)}_{el}$'s are not. Computation of the Wannier functions 
for different values of $\lambda$ is discussed in the
next section.

\subsection{Wannier Functions}
\label{wfs}
 In principle, any approach which can yield Wannier fucntions
of a crystal corresponding to its Bloch orbitals, can be used to compute 
its Born charge tensor. However, in the present work we have 
applied a framework, recently developed by us, which directly yields
the restricted-Hartree-Fock (RHF) Wannier functions of a crystalline 
insulator employing an LCAO 
approach.~\cite{shukla1,shukla2}
In our previous work we showed that one
can obtain $M$  RHF Wannier  functions, $\{|\alpha \rangle, \; \alpha =1,M\}$ 
occupied by $2M$ electrons 
localized in the reference unit cell ${\cal C}$ by solving the 
equations~\cite{shukla1,shukla2} 
\begin{equation}
( T + U
 +   \sum_{\beta} (2 J_{\beta}-  K_{\beta})   
+\sum_{k \in{\cal N}} \sum_{\gamma} \lambda_{\gamma}^{k} 
|\gamma({\bf R}_{k})\rangle
\langle\gamma({\bf R}_{k})| ) |\alpha\rangle
 = \epsilon_{\alpha} |\alpha\rangle
\mbox{,}
\label{eq-hff1}         
\end{equation}  
where $T$ represents the kinetic-energy operator, $U$ represents
the interaction of the electrons of ${\cal C}$  with the nuclei
of the whole of the crystal, while $J_{\beta}$, $K_{\beta}$,  
respectively, represent the Coulomb and exchange interactions felt
by the electrons occupying the $\beta$-th Wannier function   
of ${\cal C}$, due to the rest of the electrons of the infinite system.
The first three terms of Eq.(\ref{eq-hff1}) constitute the canonical Hartree-Fock
operator, while the last term is a projection
operator which makes the orbitals localized in ${\cal C}$ orthogonal to those 
localized in the unit cells in the immediate neighborhood of ${\cal C}$
by means of infinitely high shift parameters $\lambda_{\gamma}^{k}$'s. These
neighborhood unit cells, whose origins are labeled by lattice vectors
${\bf R}_{k}$, are collectively referred to as  ${\cal N}$. The 
projection operators along with the shift
parameters play the role of a localizing potential in the Fock matrix, and 
once self-consistency has been achieved, the occupied eigenvectors of 
Eq.(\ref{eq-hff1})  are localized in ${\cal C}$, and are orthogonal to the 
orbitals of ${\cal N}$---thus making them Wannier 
functions~\cite{shukla1,shukla2}. As far as the
orthogonality of the orbitals of ${\cal C}$ to those contained in unit cells
beyond ${\cal N}$ is concerned, it should be automatic for systems with
a band gap once  ${\cal N}$ has been chosen to be large enough. 
As in our previous calculations performed on three-dimensional ionic 
insulators,~\cite{shukla1,shukla2,albrecht,shukla3} we included  up to 
third-nearest-neighbor unit cells in the
region ${\cal N}$. 

For computing the Born charges,  first ${\bf P}^{(0)}_{el}$ is computed from
Eq. (\ref{p-wan}) using the Wannier functions of the unit cell, with all
the sublattices of the crystal in their original position corresponding
to the case $\lambda=0$. Next, the $i$-th
sublattice is displaced in the Cartesian direction $\beta$ by a small amount 
$\Delta u_{i \beta}$, and ${\bf P}^{(1)}_{el}$ is computed in a manner
identical to the previous case, except that the Wannier functions
used for the purpose are recomputed for the transformed lattice. Now that
we can compute $\Delta {\bf P}_{el}$ (cf. Eq. (\ref{delp-f})),  and compute the Born effective charge tensor by substituting it in Eq. (\ref{zstar}).

The Wannier functions obtained by solving Eq. (\ref{eq-hff1}) are 
canonical Hartree-Fock
solutions for the unit cell ${\cal C}$, and thus will satisfy the spatial
symmetries of the unit cell. In appearance they look identical to the
molecular orbitals encountered in any quantum-chemical calculation on
a finite system,
as was discussed in our earlier work.~\cite{shukla3} Therefore, by comparing
the spatial appearances of the Wannier functions for the most 
symmetric case ($\lambda = 0$), to the broken symmetry one ($\lambda=1$),
we can obtain a pictorial representation of the polarization process.

\section{RESULTS AND DISCUSSION}
\label{results}
In this section we present the results of our calculations of the 
Born effective charges for LiH, LiF, LiCl, NaF, and
NaCl. Because of the cubic nature of the underlying Bravais lattices, the
Born charge tensor for these systems has only one independent component.
In all the cases, we assumed the corresponding experimental 
fcc crystal structure, with the anion at $(0,0,0)$ position 
and the cation at $(a/2,0,0)$ position, $a$ being the lattice constant.
 The lattice constants used in the calculations
were the theoretical ones, obtained by minimizing the total energy
per unit cell at the Hartree-Fock level.

The Wannier functions used in the approach were obtained by performing
all-electron HF calculations using  
a computer program developed by us recently.~\cite{prog-wan} In order to
evaluate the centers of the Wannier functions needed for computing the 
polarization properties, we added a small subroutine to the existing module. 
The program
is implemented within an LCAO approach, employing Gaussian lobe-type
functions.~\cite{shukla2} Lobe-type functions simulate the Cartesian $p$ and 
higher angular momentum orbitals located on a given atomic site, as linear 
combinations of $s$-type functions slightly displaced from the 
site.~\cite{lobe} Because of this reason,
it is possible, that in our approach we obtain somewhat different numerical 
values of the Wannier function centers, as compared to the ones computed 
by equivalent genuine Cartesian basis functions as implemented, e.g., in
the CRYSTAL95 program~\cite{dall}.
For these calculations
we used the lobe representation of the state-of-the-art contracted Gaussian 
basis sets developed by the Torino 
group.~\cite{lih-dovesi,lif-dovesi} 
For LiH, the details of the basis set can be obtained in
ref.~\cite{lih-dovesi}, while for the alkali halides, they
are available in ref.~\cite{lif-dovesi}. 
Using these basis sets we had studied NaCl~\cite{albrecht} earlier at the 
Hartree-Fock level, therefore,
optimized lattice constants were already available for them. However,
for the remaining systems, we performed fresh Hartree-Fock calculations
to obtain the optimized lattice constants.
The theoretical lattice constant finally used
in these calculations  were 4.106 $\AA$ (LiH), 
4.018 $\AA$ (LiF), 4.633 $\AA$ (NaF), 5.262 $\AA$ (LiCl), and 
5.785 $\AA$ (NaCl).
These are in close agreement with the values 4.102 $\AA$ 
(LiH)~\cite{lih-dovesi},
4.02 $\AA$ (LiF)~\cite{lif-dovesi}, 4.63 $\AA$ (NaF)~\cite{lif-dovesi},
5.28 (LiCl)~\cite{lif-dovesi}, and 5.80 $\AA$ (NaCl)~\cite{lif-dovesi}   
reported by the Torino group.

 The computed Born effective charges  are presented in table \ref{tab-bec}.
These results were obtained by translating the sublattices of a given
crystal by the amount $\Delta u= 0.01a$ in the (100) direction.
However,
in order to ensure the stability of the results, several calculations
were performed with different directions and magnitudes of $\Delta u$,
and no significant changes in the results were observed. It was also
verified by explicit calculations that the sum total of all the
effective charges corresponding to the different atoms of a unit cell
was always zero, in agreement with its electrical neutrality. All these
reasons give us confidence as to the correctness of our results. 
 
\begin{figure}
\psfig{file=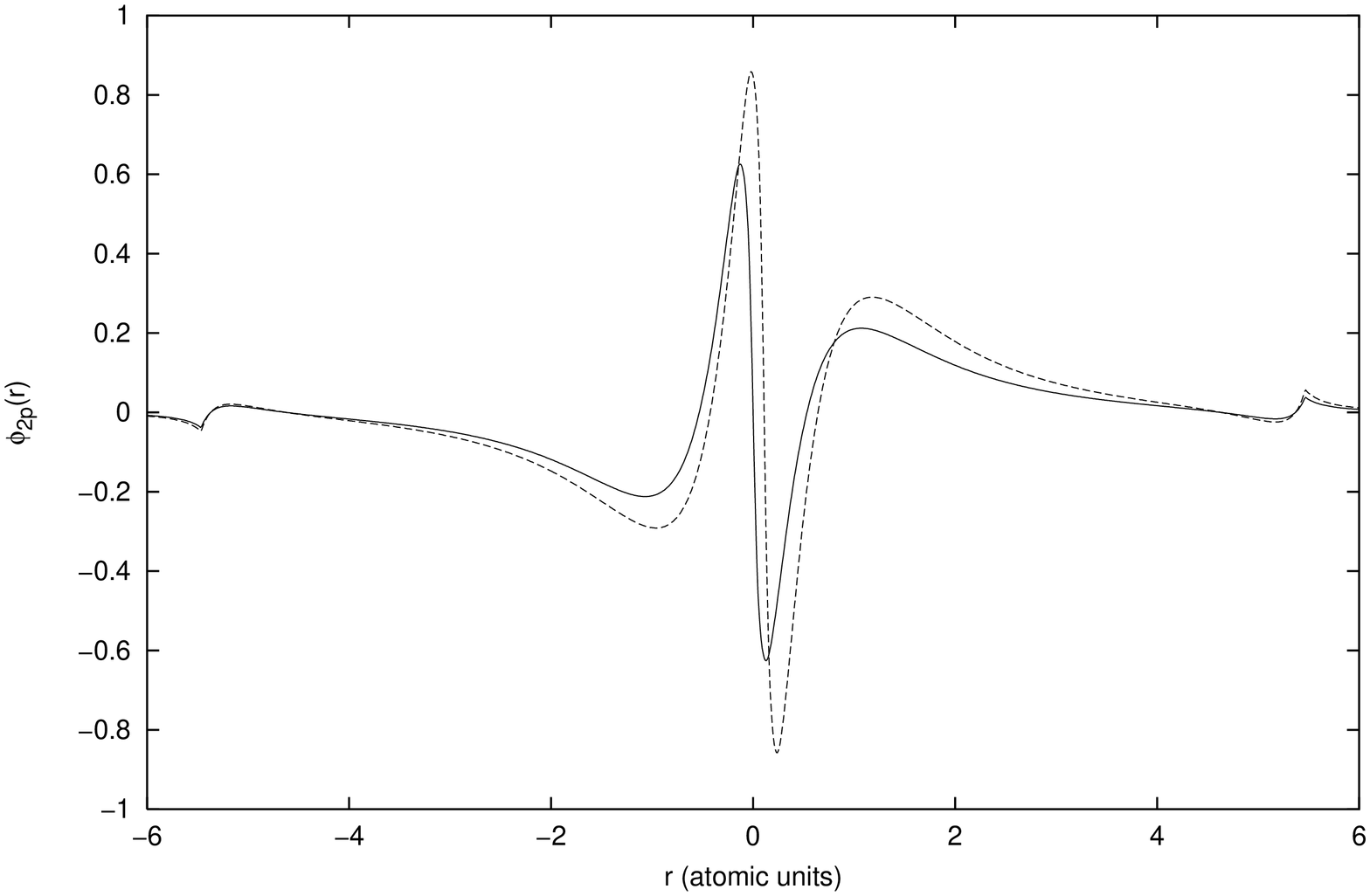,width=12.0cm,angle=-90}
\caption{Wannier functions corresponding to one of the $3p$ valence orbitals
of Cl$^-$ in NaCl before (solid line) and after (dashed line) the Cl
sublattice translation, plotted along the (100) direction.  The 
deformed lattice was obtained by translating the Cl sublattice by $0.01a$ in 
(100) direction.
Unlike the core orbitals (see Fig. \ref{fig-1s}), the valence Wannier 
function translates with significant nonrigid character, and shows 
signatures of broken symmetry.
}
\label{fig-3p} 
\end{figure}

From table \ref{tab-bec} it is obvious that the theoretical Born effective
charges obtained for LiH and the alkali halides are quite close to their
nominal ionicities. This is in perfect agreement with the intuitive picture
of these systems being highly ionic in nature.
As far as the comparison of the HF results with the experimental results 
is concerned, it is very good for LiH, LiF, and NaF. However, for NaCl
and LiCl, the disagreement is more than 10\%. 
Similar differences with respect to the experiments were
also observed by Yaschenko et al.~\cite{yaschenko} who computed the
HF Born charge of MgO to be 1.808, while the experimental value for
that compound is in the range 1.96---2.02.~\cite{yaschenko}
One possible reason for the discrepancy between the theoretical and
the experimental values of the Born charges could be the missing 
many-body effects. A qualitative discussion of these many-body effects
was given by Harrison, in the context of his  
``ion-softening theory''.~\cite{harrison-b}
When, e.g., the anionic sublattice of an alkali halide is translated, the 
bulk of the contribution to the Born charge---which, for the HF case, 
we call the mean-field 
contribution---is due to the electron transfer along the direction of the 
movement of the anion, and is associated with top-most occupied  
$p$-type Wannier function, as is obvious from Fig. \ref{fig-3p}.
However, according to Harrison~\cite{harrison-b}, because of the 
many-body effects, we can have a single (virtual) excitation from the top 
$p$-type occupied Wannier function (the bonding orbital) into the  first 
unoccupied Wannier function (the antibonding orbital) on the nearest-neighbor 
cations, thereby, modifying the Born charge. This virtual charge fluctuation, 
in effect, introduces some covalency
into the system as compared to the mean-field HF results. In its simple
parametrized form, the ion-softening theory of Harrison predicts a uniform
value of $Z^*=1.16$, for the alkali halides.~\cite{harrison-b} This
value of $Z^*$, although reasonable, is clearly at variance with
the experimental results which show a clear variation in the
$Z^*$ values of different alkali halides. Therefore, it is of interest
to borrow the essence of the many-body effects incorporated in the
ion-softening theory, and apply it to these systems within a rigorous 
{\em ab initio} formalism, to test its applicability. Indeed, this is what
we intend to explore in a future paper. 

 In table \ref{tab-wan} we give the detailed contributions of various
Wannier functions to the Born effective charges of the
alkali halides, when the anionic sublattice is translated. It is clear 
from the table that the low-lying core-like orbitals basically translate
rigidly along with the nuclei. Nonrigid translation is seen mainly
for the $n$s and $n$p Wannier functions of the anion, where $n$ defines
the top of the valence band. In particular, $n$s orbital gains some
effective charge at the expense of the $n$p orbital. The case of NaF is
an exception to this rule where the Na 2p Wannier function makes a significant
contribution to the effective charge (-0.216). However, this contribution
is due to an accidental near degeneracy of the sodium 2p Wannier function 
with the 2s Wannier function of fluorine, which leads to their mixture 
when the HF equations 
(cf. Eq. (\ref{eq-hff1})) are solved. Because of this reason,    
some of the Born 
effective charge associated with the 2s Wannier function of F is 
transferred to the 2p function of Na (cf. table \ref{tab-wan}). However, 
this is an instructive example
of the nonuniqueness of the individual Wannier functions. But, as should be
the case, the total Born charge of fluorine in NaF is free of this 
ambiguity associated with the individual Wannier functions, in that it has a 
normal value of -0.956. 

\begin{figure}
\psfig{file=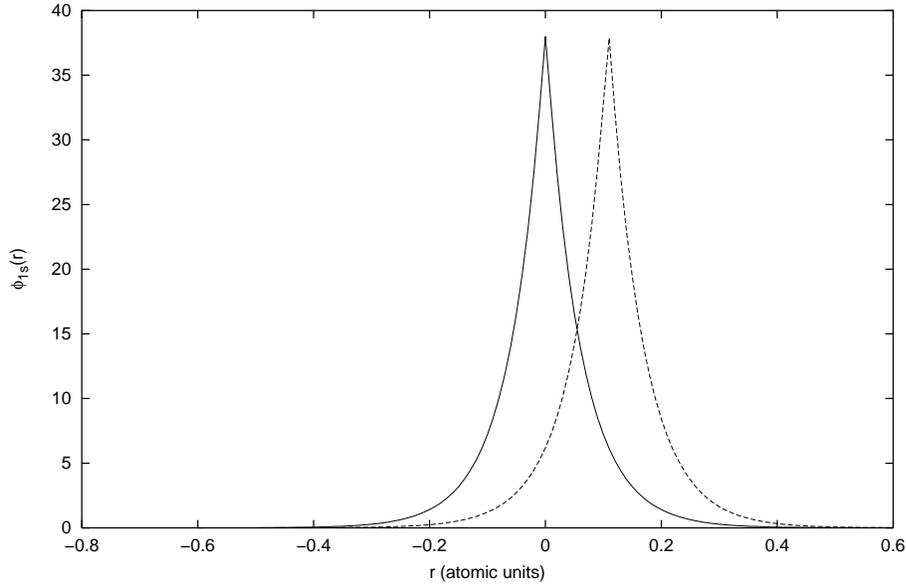,width=12.0cm,angle=-90}
\caption{Wannier functions corresponding to the 1s core orbital
of Cl$^-$ in NaCl for the undeformed lattice (solid line), and the
deformed lattice (dashed line).
Rest of the information is
same as in the caption of Fig. \protect\ref{fig-3p}. 
 Clearly,
as expected, the 1s Wannier function translates rigidly with the sublattice.}
\label{fig-1s} 
\end{figure}

It is also instructive to examine
the polarization process pictorially, as depicted by Wannier functions.
We shall do so for the specific case of NaCl. The 1s and 3p Wannier 
functions, localized on the Cl$^-$ site of the unit cell, are plotted along
the (100) direction, in Figs. \ref{fig-1s} and 
\ref{fig-3p}, respectively, both before, and after, the translation of
the Cl sublattice. 
 As discussed earlier, on intuitive grounds we would expect
the highly localized 1s Wannier function, which is the deepest-lying core
orbital, to move rigidly with the
nucleus. On the other extreme, we would expect the 3p Wannier function, which
forms the top of the valence band, to show significant nonrigid behavior,
because of its relatively delocalized character. This is indeed what
we observe in Figs.  \ref{fig-1s} and \ref{fig-3p}, respectively.
Owing to the perfectly cubically-symmetric crystal field that the Cl site 
sees in the undeformed lattice, one would expect the corresponding 
3p Wannier function to exhibit perfect antisymmetry about its center. 
Once the Cl sublattice is moved along the
(100) direction, the crystal symmetry is reduced, and one would expect
to see the signatures of the broken symmetry in the 3p Wannier function
of Cl. Both these phenomenon are clearly visible in Fig. \ref{fig-3p},
where, for the undeformed lattice the 3p Wannier function is perfectly
antisymmetric about its center, while for the deformed case, it is
no longer so, and it shows clear signs of induced polarization due to
broken symmetry.

\section{CONCLUSIONS}
\label{conclusion}

In conclusion, we have applied the Berry-phase-based theory of
macroscopic polarization, developed by King-Smith and Vanderbilt~\cite{king},
 to obtain  the
benchmark values for the Born effective charges of several ionic compounds 
at the Hartree-Fock level. In the present work, we have utilized the
Wannier functions as the single particle orbitals, and demonstrated
that they lead to a pictorial description of the polarization process.
As far as our results are concerned, they are in good agreement with
the experiments for all the systems except LiCl and NaCl, where the
disagreement with the experiments was more than 10\%. One of the
reasons behind this disagreement could be that the many-body effects
in these systems are significant. 
Although, there have been generalizations
of the theory of macroscopic polarization to include many-body 
effects,~\cite{ortiz} their implementation is not as straightforward
as the single particle theory. 
Recently, we have generalized our Wannier-function-based approach
to include many-body effects by systematically enlarging
the many-particle ground-state wave function by considering virtual
excitations from the space of the occupied Wannier functions to that
of the virtual ones.~\cite{shukla7} The approach was demonstrated by
computing the correlation contributions to the total energy per unit
cell of bulk LiH~\cite{shukla7}. 
In a future paper, we intend to generalize our approach to
compute the influence of many-body effects on macroscopic polarization
properties as well.

\acknowledgements
I am thankful to Professor R. Resta for clarifying the evaluation
of the experimental value of the Born effective charge of MgO
reported in an earlier work of his(ref.~\cite{poster2}).

\begin{table}  
 \protect\caption{Born effective charges of different ionic
crystals obtained in this work, as compared to the experimental values.}
 \protect\begin{center}  
  \begin{tabular}{lccc} \hline
 \multicolumn{1}{c}{Crystal} & \multicolumn{2}{c}{Born Effective Charge} 
& \multicolumn{1}{c}{Percentage Error} \\ 
      & This Work  & Experiment   &  \\
      & $Z^*(theory)$ & $Z^*(exp)$ &
$\frac{(Z^*(theory)-Z^*(exp))}{Z^*(exp)}\times 100$ \\ \hline \hline 
LiH     & 1.046      & 0.991$\pm$0.04$^a$ & 5.5       \\
LiF     & 0.998      & 1.045$^b$      & -4.5   \\
LiCl    & 1.036      & 1.231$^b$      & -15.6 \\
NaF     & 0.956      & 1.024$^b$      & -6.6  \\
NaCl    & 0.985      & 1.099$^b$    & -10.4  \\
\end{tabular}                      
   \end{center}  
$^a$ Obtained from experimental values of the Szigeti charge $Z_s$, and the
high-frequency dielectric constant $\epsilon_{\infty}$, using
the relation $Z^{*}= \frac{(\epsilon_{\infty} + 2)}{3} 
Z_s$.~\cite{lucovsky}
Experimental $Z_s$ and $\epsilon_{\infty}$ were reported in 
Ref.~\cite{exp-lih}\\
$^b$  Obtained from the experimental values of $Z_s$ and $\epsilon_{\infty}$
reported in ref.~\cite{exp-ah} \\
\label{tab-bec}    
\end{table}  
\begin{table}  
 \protect\caption{The contribution  of individual Wannier functions to the
Born effective charge of the alkali halides when the 
anion($A$) sublattice with nuclear charge $Z_{nuc}$  was 
translated, holding the cation ($C$) sublattice(s) fixed. Nominal
ionicity of the anion is given in the parenthesis right below its 
Born effective charge.}
 \protect\begin{center}  
  \begin{tabular}{cccccc} \hline
 \multicolumn{1}{c}{Wannier function} & {Nominal Charge } & 
        \multicolumn{4}{c}{Born Effective Charge} \\ 
        &       & LiF   & LiCl  & NaF   & NaCl \\ \hline
1s ($C$)&0.000 & 0.001 & 0.002 & 0.000 &  0.000 \\
2s ($C$)&0.000 & ---   & ---   &-0.007 & -0.001  \\
2p ($C$)&0.000  & ---   & ---   &-0.216 &  0.002  \\  
1s ($A$)&-2.000 &-2.000 &-2.000 &-2.000 & -2.000 \\
2s ($A$)&-2.000 &-2.061 &-2.002 &-1.830 & -2.001 \\
2p ($A$)&-6.000 &-5.938 &-5.998 &-5.903 & -5.999 \\
3s ($A$)&-2.000 & ---   &-2.107 & ---   & -2.081 \\
3p ($A$)&-6.000 & ---   &-5.931 & ---   & -5.905 \\ \hline
$Z_{nuc}$ &       & 9.000 &17.000 & 9.000 & 17.000 \\
Total   &       & -0.998  & -1.036  & -0.956 &-0.985  \\
       &       & (-1.000) & (-1.000) & (-1.000) & (-1.000) \\ \hline
\end{tabular}                      
\end{center}  
\label{tab-wan}    
\end{table}  
\end{document}